\documentclass[dvips,, 12pt]{article}
\usepackage{latexsym, epsfig, amsmath, graphicx, amssymb}
\usepackage{mathrsfs, amsfonts, graphics}
\begin{document}

\begin{center}
\textbf{\Large ON QUANTUM EFFECTS NEAR A BLACK HOLE SINGULARITY}\\[1.2cm]
Asghar Qadir$^{1}$ and Azad A. Siddiqui$^{2}$\\[3ex]
$^{1}$Inernational Centre for Theoretical Physics, Trieste, Italy
\\[0pt]
$^{2}$Mathematics Department, Quaid-i-Azam University, Islamabad, Pakistan,\\[0pt]
E-mail: azad@ceme.edu.pk \\[0pt]

\bigskip

\bigskip

\bigskip

\textbf{Abstract}
\end{center}

\begin{quotation}
It is pointed out that the claim made by Joshi and Joshi [1], has
not been rigorously demonstrated by them. A simpler and more
correct proof is provided.
\end{quotation}

\section{References}

1. P.S. Joshi and S.S. Joshi, Class. Quantum Grav. 5 (1988)
L191.\newline 2. V.A. Belinski, I.M. Khalatnikov and E.M.
Lifshitz, Adv. in Phys. lj> (1970) 525; Sov. Phys. JETP 33 (1971)
1061; 35 (1971) 838. \newline 3. C.W. Misner, K.S. Thorne and J.A.
Wheeler, Gravitation (W.H. Freeman, 1973). \newline 4. S.W.
awking, Astrophysical Cosmology: Proc. Study Week on Cosmology and
Fundamental Physics, Eds, H.A. Bruck, G.V. Coyne and M.S. Longair
(Pontificae Academica Scripta Varia 413 (1982) 563): Relativity,
Groups and Topology II.E'ds. B.S. DeWitt and R. Stora
(North-Holland, 1984); J.B. Hartle and S.W. Hawking, Phys. Rev.
D28 (1983) 2960.

\end{document}